\documentclass[11pt]{elsart}

\usepackage{graphicx}
\usepackage{epsf}
\usepackage{wrapfig}
\usepackage{here}

\begin{document}

\begin{frontmatter}

\title{Observations of the Crab Nebula with the HEGRA system 
  of IACTs in convergent mode using a topological trigger}

\thanks[corresp]{Corresponding author. \\ E-mail address: lucarel@gae.ucm.es}
\author[1,2]{F. Lucarelli\thanksref{corresp}}, 
\author[1]{A. Konopelko},
\author[1]{F. Aharonian},
\author[1]{W. Hofmann},
\author[1]{A. Kohnle},
\author[1]{H. Lampeitl}, 
\author[2]{V. Fonseca}

\bigskip

\address[1]
{Max-Planck-Institut f\"ur Kernphysik, D-69029 
Heidelberg, Germany}
\address[2]
{Facultad de Ciencias F\'{\i}sicas, Universidad Complutense,
E-28040 Madrid, Spain}

\begin{abstract}
Routine observations of the Crab Nebula for a total of about 250~hrs, 
performed with the HEGRA stereoscopic system of 5 imaging atmospheric 
\v{C}erenkov telescopes in the standard o\-pe\-ra\-tio\-nal mode, have 
proven the energy threshold of the system to be 500~GeV for small 
zenith angles ($\theta \leq 20^\circ$). A {\it topological trigger} 
applied along with the {\it convergent} observational mode allows to 
reduce noticeably the energy threshold of the system down to 350~GeV. 
Here we present the relevant Monte Carlo si\-mu\-la\-tions as well 
as the analysis results of 15~hrs Crab Nebula data taken in such an 
observational mode. From the Crab Ne\-bu\-la data, the final energy 
threshold was found to be 350~GeV. The estimated $\gamma$-ray flux 
from the Crab Nebula above 350~GeV is 
$(8.1\pm 0.1_{stat}\pm0.2_{syst})\times 10^{-11}\,\,
\rm ph\,cm^{-2}\,s^{-1}$, which is consistent with recent measurements
reported by the STACEE, CELESTE, CAT, and Whipple groups.

\vspace{3mm}


\end{abstract}
\end{frontmatter}

\section{Introduction}
Ground-based TeV $\gamma$-ray astronomy has effectively ex\-ploi\-ted
i\-ma\-ging at\-mo\-sphe\-ric \v{C}e\-ren\-kov te\-le\-sco\-pes (IACT) 
in the study of $\gamma$-ray emission arising from a few well established 
sources, as well as for a general search amongst large numbers of 
$\gamma$-ray source candidates \cite{cw99}. Given the rather steep energy 
spectra measured for dis\-co\-ver\-ed $\gamma$-ray sources as well 
as for those pre\-dic\-ted as potential can\-di\-da\-tes, a
substantial advancement in the sen\-si\-ti\-vi\-ty of IACT techniques can be reached 
if con\-si\-de\-ra\-ble lowering of the energy threshold\footnote{The position 
of a peak in the differential $\gamma$-ray detection rate, $E$ [TeV], is
usually assumed as the effective energy threshold of the instrument.} 
of the instrument is achieved. For such spectra the low 
energy threshold provides a high $\gamma$-ray rate, which gives a
high sensitivity of the instrument. Nowadays a number of instruments 
perform observations at the effective energy threshold as low as 
250~GeV \cite{cw99}. 

Further reduction of the e\-ner\-gy thre\-shold could be a\-chie\-ved by using of {\it (i)} larger optical 
reflectors (10-20~m in dia\-me\-ter) to increase the photon collection 
efficiency for the low energy $\gamma$-ray showers; {\it (ii)} imaging 
cameras equipped with pixels of a relatively small angular size 
($0.1^\circ - 0.15^\circ$), to reduce the background night sky light per
pixel; and finally, {\it (iii)} by operating a number of telescopes simultaneously in the 
{\it stereoscopic} observational mode, which allows to reduce the 
trigger threshold by requiring coincidences in a number of 
telescopes. 

The latter approach was proven by the system of 5 imaging 
atmospheric \v{C}erenkov telescopes built by the HEGRA ({\it High Energy
Gamma-Ray Astronomy}) collaboration at La Palma, Canary Islands.  
Even for rather small optical reflectors of 8.5$\rm \,\,m^2$ for each 
of the telescopes and relatively modest angular size of the ca\-me\-ra 
pixels of 0.25$^\circ$ the energy threshold of the system is about 
500~GeV using the nominal observational mode. The corresponding 
effective dynamic energy range of the $\gamma$-ray observations with 
the HEGRA system of IACTs extends from 500~GeV up to 20~TeV, as shown 
by the observations of the well established TeV $\gamma$-ray source~- 
the Crab Nebula \cite{me00}. The extended HEGRA data sample for the 
Crab Nebula permitted, thanks to the good energy resolution of the HEGRA
system of about 10-20\%, measurements of the Crab Nebula spectrum with 
high accuracy over the entire dynamic energy range. 

Lowering the e\-ner\-gy thre\-shold of the \v{C}e\-ren\-kov 
te\-le\-sco\-pes is not only motivated by the enhancement of the 
instrument sensitivity but it is also physically important to extend 
the spectral studies of the $\gamma$-ray emission to lower energies. 
For example, inverse Compton (IC) modeling of the $\gamma$-ray 
spectrum of the Crab Nebula, taken along with the EGRET detection at 
GeV energies \cite{aa96,dj96}, predicts a substantial spectral 
flattening down to energies of 200-300~GeV, whereas above 1~TeV the 
spectrum has rather a power-law shape of a spectral index of -2.6. The 
measurement of such change in the spectrum slope would favor the IC 
model of $\gamma$-ray emission. First evidence for deviation from a 
straightforward power law was suggested by the WHIPPLE group
(see~\cite{hillas2,krenn}) and, subsequently, it was confirmed by other 
measurements made within that energy range by STACEE \cite{o01}, 
CELESTE\cite{n01}, and CAT \cite{pir01}. 
 
In this paper we discuss a new approach to detect low energy $\gamma$-rays with 
the system of HEGRA IACTs by use of a {\it topological} system trigger 
along with the {\it convergent} observational mode. It allows to 
reduce considerably the trigger threshold and cor\-re\-spon\-din\-gly 
the e\-ner\-gy thre\-shold down to 300-350~GeV. Relevant Monte Carlo 
simulations (see Section~III) and trigger tests (see Section~IV) have 
been done in order to prove the efficiency of such observational 
technique. Finally, we performed 15~hrs observations of the
Crab Nebula using the topological trigger technique with the
convergent observational mode. Analysis of these data has shown an 
excess attributed to $\gamma$-ray showers at energies well below
500~GeV (Section~V-VI). Assuming the shape of the energy spectrum as 
measured by HEGRA \cite{me00} we estimated the integral flux of
$\gamma$-rays from the Crab~Nebula above 350~GeV (Section~VII).

\section{Trigger modes}
For most of its operational time the HEGRA IACT system was running 
using a {\it standard two-level} trigger~\cite{b98}. For that, the
system telescopes were triggered {\it locally} when the signals in at 
least two photomultipliers (PMTs) exceeded a given trigger threshold
of 7-8~ph.e.\footnote{Hereafter ph.e. stands for photoelectrons.} 
Moreover, those two PMTs should be neighbors (2NN, two-nextneighbour) 
amongst the 271 PMTs which form the camera. That permits a 
reduction of the accidental background triggers, caused by the light
of the night sky, by a factor of 48. For the second level trigger a 
{\it global} (system) trigger requires at least two telescopes being 
triggered by the same shower within a given coincidence time window. The 
telescopes CT2, CT4, CT5, and CT6 are $\simeq 80 \,\rm m$ apart from 
the central telescope CT3\footnote{The first telescope (prototype) 
built by the HEGRA collaboration and numbered CT1, it is not included 
in the stereoscopic system.} and the inter-telescope coincidence 
window is set to 70~ns. Detailed comparisons of the actual hardware 
event rate with the si\-mu\-la\-tions for different trigger configurations 
and thresholds are given in \cite{me99}. 

A trigger threshold of 7-8~ph.e. is high enough to suppress accidental 
triggers due to illumination of the ca\-me\-ra pixels by background 
night sky light. At first, the choice of the trig\-ger is
de\-ter\-mined by the a\-ve\-ra\-ge a\-mount of night sky light 
pho\-tons hit\-ting the ca\-me\-ra PMT (pixel) with\-in a trigger
gate, which is of 20-30~ns. This value is e\-qui\-va\-lent to 1~ph.e. for the 
HEGRA telescopes. It is set by the size of the optical reflector, the 
angular size of the camera pixels, and the efficiency of the light
detectors (PMTs). Secondly, choice of the trigger threshold depends on 
the designed trigger logics and finally on the total number of
ca\-me\-ra pixels included into the trigger zone. For a smaller number 
of pixels in the trigger zone, a reduced number of accidental
coincidences is possible, and consequently lower trigger thresholds
can be chosen. 

Trigger threshold limits the energy of the atmospheric showers, which
should be sufficiently powerful to generate a certain number of 
\v{C}erenkov photons and trigger the camera. Note that low energy 
$\gamma$-ray-induced atmospheric showers can produce a sufficient
amount of \v{C}erenkov light photons only at rather small distances
from the shower core ($r\leq50\,\,\rm m$). Using only 4-telescope
coincidences for the HEGRA system of IACTs, these low energy
$\gamma$-ray showers are concentrated within the geometrical
area of the array. In addition, the \v{C}erenkov light images of such
showers occur in specific parts of the camera for each of the system  
telescopes, which are determined by the location of the telescopes
within the array. By intention, to detect only low energy $\gamma$-ray 
showers, one can restrict the trigger zone to some specific part of
the camera, which is sensitive to these events. A reduction of the
trigger zone by a factor of about 7 allows to decrease the value of
the trigger threshold down to 4~ph.e. and thus achieve a considerable
enhancement in the trigger rate of low energy $\gamma$-ray showers. 
This so-called {\it topological} trigger mode is effective
for detection of low energy $\gamma$-rays, whereas the rate of the
high energy $\gamma$-rays well above the energy threshold of the
system, will be drastically reduced due to rather narrow trigger
zone.

One could try to operate the HEGRA telescopes in standard trigger mode 
at the very low trigger threshold of 4~ph.e. However, 
that would imply a drastic increase in the accidental trigger rate of up to 100~Hz and 
an increase of up to a 30\% in the dead time (against a rate of 15~Hz and 5\% 
dead time for the nominal trigger threshold of 7~ph.e.(see~\cite{b98})). 
Moreover, current data acquisition system does not support such rate. 
It would be possible then, to increase the trigger multiplicity and 
ask for at least 4 telescopes out of 5 to be triggered, but in this 
case, as we already mentioned previously and we show below through 
the Monte Carlo simulations, we do not need to use the whole camera, 
since only certain parts of the camera will be covered by the images 
of low energy $\gamma$-rays. 
  
In the next sections we discuss in detail the implications of using 
a {\it topological} trigger mode.

\section{Simulations}
Based on the HEGRA Monte Carlo simulations, we have estimated the
detection rates of cosmic rays and $\gamma$-rays, as well as the
energy threshold of $\gamma$-rays for the HEGRA IACT array using 
a {\it topological} trigger mode. For such trigger mode one 
has to use restricted trigger zones within the camera field of view
for each of the system telescopes. These zones are adjusted to the 
regions of most frequent appearance of the $\gamma$-ray images. 
For instance, the restriction on impact distance from the center of 
the IACT system at 50~m yields a specific distribution of the 
$\gamma$-ray events in the telescope camera focal plane (see 
Figure~1). Note that the maximum impact distance for the simulated 
$\gamma$-ray showers was 500~m. As it is seen from Figure~1 all 
events are localized within an angular area roughly corresponding 
to 19-37 camera pixels. The location of the trigger zone in the 
camera field of view differs from one telescope to another and is 
determined by the geo\-me\-tri\-cal layout of the telescopes in the 
array. In simulations we have assumed that the optical axes of all 
system telescopes are parallel, which represents the conventional 
mode of a telescope tracking. One can see in Figure~1 that the 
corresponding cluster of the triggered events is shifted by about 
$0.6^\circ$ from the camera center, which roughly corresponds to 3 
camera pixels. 

% ----- Figure 1 -------------------------------------------------------------
\begin{figure}[htp]
\begin{center}
\includegraphics[width=0.6\linewidth]{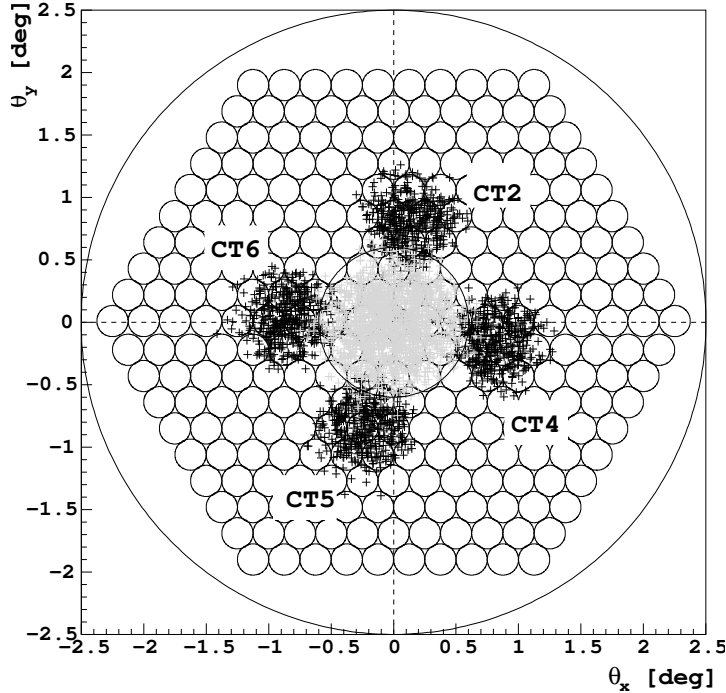}
\end{center}
\caption{\em Two-dimensional distribution of the centroid positions for $\gamma$-ray
images, triggering at least 4 telescopes of the HEGRA IACT array
and passing the selection of the impact distance from the center of the
system less than 50~m (black crosses). Simulations have been done for showers
at zenith. Data for CT3 are not shown. In convergent mode the centers of gravity are shifted towards
the ca\-me\-ra center (grey crosses).}
\label{fig1}
\end{figure}

% ----------------------------------------------------------------------------
 
We have made calculations of the cosmic and $\gamma$-ray detection
rates using two different trigger modes: a {\it standard} local trigger condition of $\rm 2NN/271>q_0$, 
where $\rm q_0$ is the trigger threshold measured in photoelectrons, and 
a {\it topological} trigger mode of $\rm 2NN/19>q_0$, where only 19~pixels were selected for each telescope 
according to the spe\-ci\-fi\-cal\-ly established trigger zone. 
In both cases, a global trigger mode defined as at least 4 telescopes out of 5 to be triggered by a shower
was applied. The detection rates of cosmic rays and $\gamma$-rays calculated for different values
of trigger thresholds ($\rm q_0$) are shown in Figure~2. 

% ----- Figure 2 -------------------------------------------------------------
\begin{figure}[htbp]
\begin{center}
\includegraphics[width=0.55\linewidth]{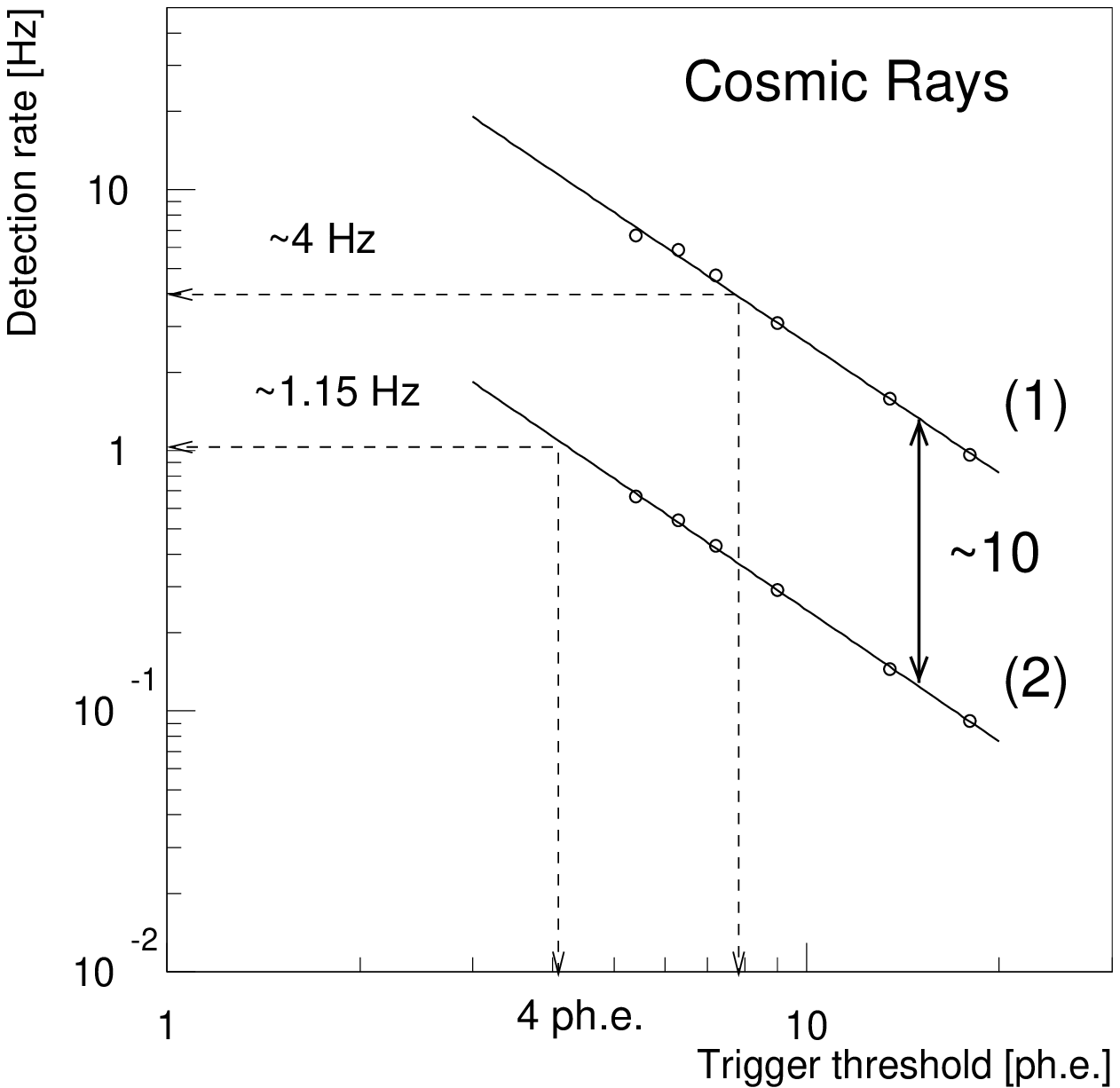} 
\hspace{5.mm}
\includegraphics[width=0.55\linewidth]{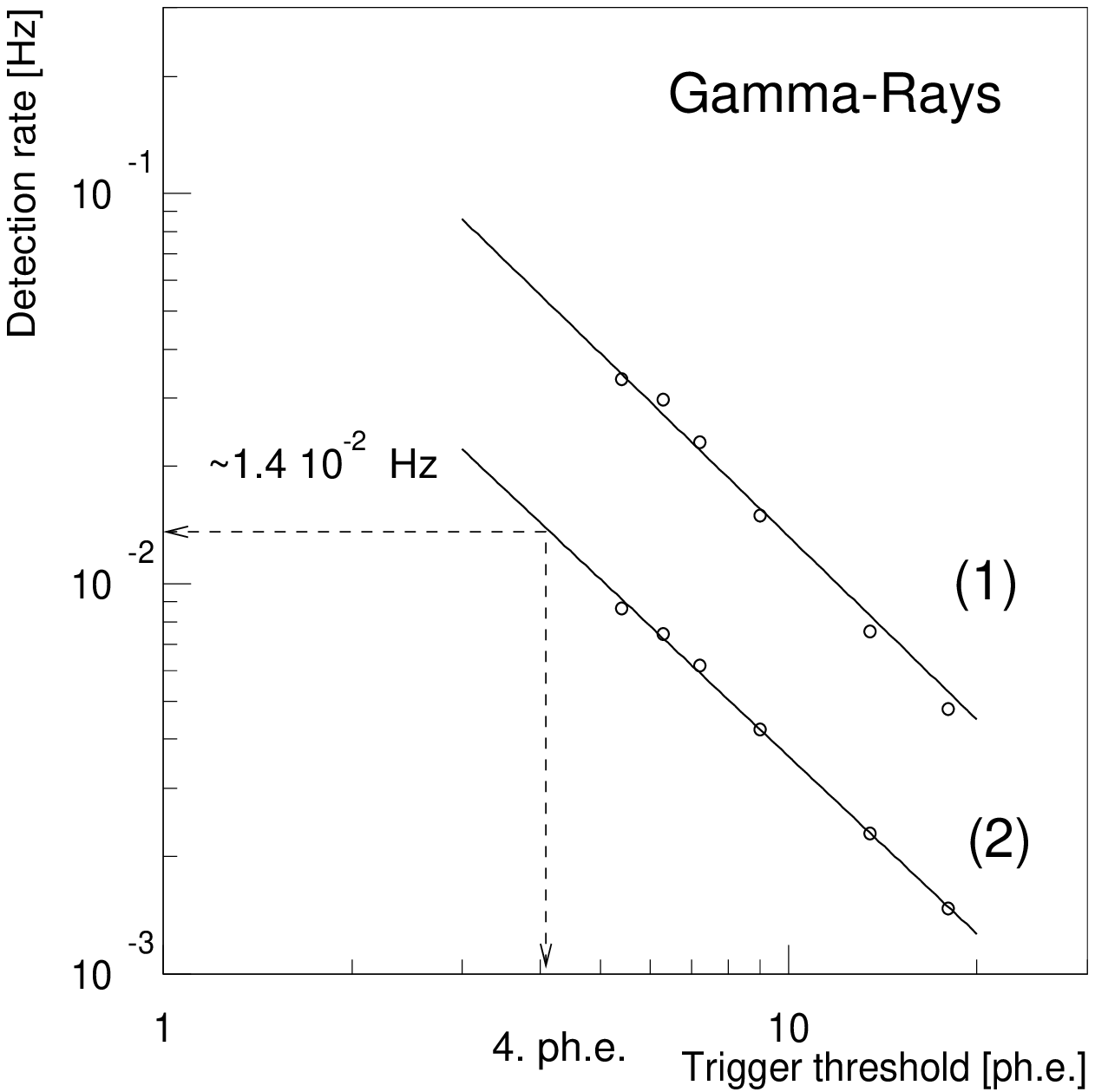}
\end{center}
\caption{\em Detection rates of cosmic rays and $\gamma$-rays 
calculated from MC simulations for different values of trigger 
thresholds ($\rm q_0$) and for two different trigger modes, (1) {\it
standard}: 4 telescopes out of 5, $\rm 2NN/271>q_0$, and (2) {\it
topological}: 4 telescopes out of 5, $\rm 2NN/19>q_0$. Also the correspondent 
fits and the rates for $\rm q_0=4~$ph.e. are shown. Simulations have been done 
for showers at the zenith using parallel tracking in both cases.}
\label{fig2}
\end{figure}
% ----------------------------------------------------------------------------

The HEGRA system of IACTs runs in the {\it standard} trigger mode 
with the trigger threshold set to $\simeq 8$~ph.e. That yields a 
corresponding detection rate of about 4~Hz for the 4-fold
coincidences, which is in a good agreement with the measurements. Note
that the detection rates of cosmic rays differ by a factor of about 10
between the two dif\-fe\-rent trigger modes (see Figure~2), roughly agreeing
with the simple estimate based on the number of the pixels included in
the trigger ($271/19 = 14$). In reality, in the experimental data we
have found a non uniform rate over the whole camera field of view,
which is due to a non equal optical smearing (it is broader at larger
angular distances from the center of FoV), restricted number of the
trigger patterns for the images being close to the camera edge (i.e.,
two pixels on the ring), etc. All these effects reduce the number of
possible combinations for the 2NN/M trigger logic (which is roughly 
$3 \times M$, where {\em M} is the total number of pixels in the camera,
for HEGRA telescopes $M=271$) and, consequently, the total rate in the 
standard trigger mode. Using the {\it topological} trigger mode with
the trigger threshold set at 4~ph.e. the cosmic ray detection rate is
about 1.15~Hz. That is somewhat higher than the measured rate for a
configuration with 19 pixels in trigger (see section~4). However,
given the imperfect adjustment of the trigger window for the rate
measurements (due to telescope pointing error) as well as slight
differences between Monte Carlo assumed trigger thresholds and those
actually used in observations (e.g. due to a 20\% reduced mirror
reflectivity etc.), one can conclude that there is a rather good agreement between
measured and calculated rates.  

For the {\it topological} trigger mode the calculated $\gamma$-ray
rate is  $~1.4 \cdot 10^{-2}$~Hz (50 $\gamma$'s per hr) (see
Figure~2). In calculations we assumed the $\gamma$-ray energy
spectrum of the Crab Nebula as measured by the HEGRA collaboration
\cite{me00}. Dependence of the evaluated energy threshold on the value
of the trigger threshold, $q_0 [ph.e.]$, can be well fitted by 
\begin{equation}
  log(E [TeV]) = -1.0 + 0.9 \cdot log(q_0) .
\end{equation}
Thus for the {\it topological} trigger mode and conventional trigger 
threshold, $\rm q_0$ = 8~ph.e., the energy threshold is about 500~GeV,
whereas for the reduced trigger thres\-hold, $\rm q_0$ = 4~ph.e., the
energy threshold is of $\simeq 350$~GeV. Such substantial
lowering of the energy thres\-hold by a factor of 1.4 with respect to the {\it
standard} trigger mode could only be achieved by implementing a corresponding
upgrade of the telescope hardware, e.g. towards much larger effective
collection area of the optical reflector. 

For the event classification, both the reconstructed shower direction  
and the averaged angular size of the \v{C}erenkov light were used in 
the present analysis. Images recorded at such low energy threshold
contain a rather small number of photoelectrons, and image orientation 
for those events is poorly determined. However, using four images per 
each individual event one can substantially improve the angular
resolution. Thus, for a rather loose angular cut of $\theta^2 \leq 0.05
\,[deg^2]$, where $\theta^2$ is the squared angular distance of the 
reconstructed shower arrival direction to the source position, the
Monte Carlo simulations predict a $\gamma$-ray acceptance (which is the
fraction of remaining events after applying the directional cut),
$\kappa_\gamma^{dir}$, of about 0.7. The predicted acceptance for the cosmic ray, 
$\kappa_{CR}^{dir}$, is of about 0.05, which corresponds to
a rejection factor for the cosmic rays of $\kappa^{-1}_{CR}\simeq
20$. Therefore, the directional information results in an enhancement 
of the signal-to-noise ratio by a factor $Q^{dir} =
\kappa_\gamma^{dir}/ \sqrt{\kappa_{CR}^{dir}}\simeq 3.2$. 

Further improvement of the signal-to-noise ratio could be achieved by 
using the differences in the image shape between the cosmic ray- and  
$\gamma$-ray-induced showers~\cite{hillas}. In general, the angular 
size of the \v{C}erenkov light images depends on the geometrical
distance of the shower axis to the telescope and the primary energy of
the shower. For a better separation of the $\gamma$-ray images out of
the cosmic ray images one has to re-scale the angular image size 
(WIDTH parameter \cite{hillas}) with respect to the impact distance
and image SIZE (total number of photoelectrons in image) \cite{me99}. 
Showers selected by the {\it topological} trigger are concentrated
within the telescope array, and therefore occur at small impact 
distances from the center of the array. The dynamic energy range of  
these events is also very limited, from 100~GeV up to 1~TeV. 
In such a case the scaling of the angular image size is not necessary,
and one can simply use the initial WIDTH parameter for an effective
image classification. All triggered cosmic ray and $\gamma$-ray events 
can be plotted versus {\it mean WIDTH} $\langle W \rangle$, which is 
calculated as:   
\begin{equation}
  \langle W \rangle = \sum_i^N A_i W_i / \sum_i^N A_i
\end{equation}
where $W_i$ and $A_i$ are WIDTH parameter and SIZE parameter,
respectively, measured for each telescope out of N (N=4 or 5) in the
system. The mean image SIZE is given by 
\begin{equation}
  \langle A \rangle = \frac{1}{N} \sum_i^N A_i \,\,.
\end{equation}

% ----- Figure 3 -------------------------------------------------------------
\begin{figure}[b]
\begin{center}
\includegraphics[width=0.6\linewidth]{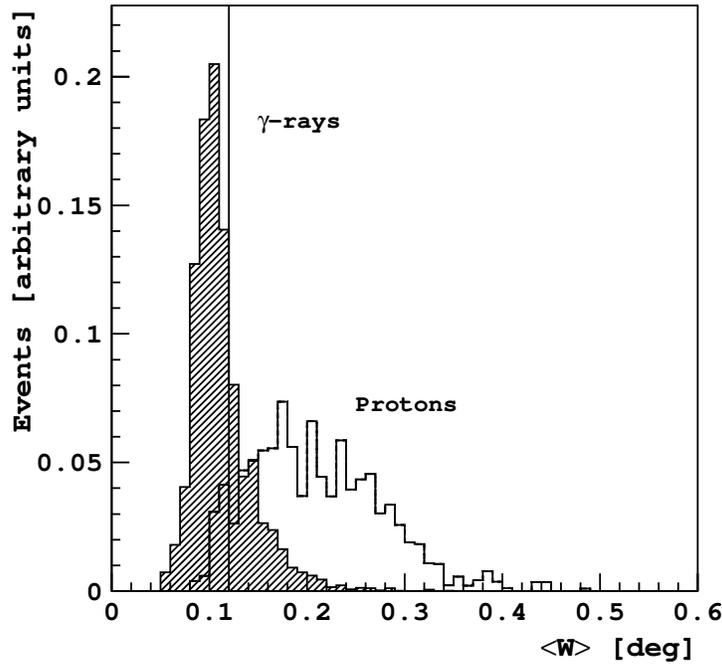}
\end{center}
\caption{\em Distributions of the parameter $\langle W \rangle$ for cosmic ray and
$\gamma$-ray showers. The vertical solid line indicates the cut applied in the 
analysis, $\langle W \rangle \le 0.12^\circ$.}
\label{fig3}
\end{figure}
% ----------------------------------------------------------------------------

Both distributions are shown in Figure~3 for cosmic ray and $\gamma$-ray events. 
One can see in Figure~3 that the parameter $\langle W \rangle$
enables a rather good rejection of cosmic ray showers. The post-cut
$\gamma$-ray and cosmic ray acceptances based on the image shape cut
are $\kappa_\gamma^{sh} \simeq 0.72$, $\kappa_{CR}^{sh} \simeq
0.082$, respectively, with a corresponding signal enhancement factor of 
$Q^{sh} = \kappa_\gamma^{sh}/\sqrt{\kappa^{sh}_{CR}} \simeq 2.5$.  

The signal-to-noise ratio in observations with the HEGRA system  of
IACTs running with the to\-po\-lo\-gi\-cal trig\-ger for the
re\-du\-ced trig\-ger thres\-hold can be finally estimated as 
\begin{equation}
  \rm S/N = R_\gamma/(R_{CR})^{1/2} Q^{dir}Q^{sh} \sqrt{t} \,\,.
\end{equation}
For 1~hr observations of the Crab Nebula, we expect a significance
$\rm S/N \simeq 6\sigma$ for a 19 PMT trigger zone. Ne\-ver\-the\-less, for
the trigger zone of 37 pixels actually used in the observations of the
Crab Nebula (see Section~V), the integral $\gamma$-ray rate from a
point-like source remains nearly the same, whereas the rate of cosmic rays,
which are randomly distributed over the solid angle, increases due to
the substantially broader trigger region. The choice of a 37 pixel trigger
zone was forced by the uncertainties in the orientational tilt of the
telescopes. Thus we may expect a lower significance of around
4$\sigma/\sqrt{hr}$. Note that the expected $\gamma$-ray rate was 
calculated assuming the Crab Nebula spectrum at energies above 
100~GeV as a power-law with a spectral index of 2.6 and the 
flux normalization taken from \cite{me00}. Actually, as
measured by the HEGRA IACTs system~\cite{me00}, the Crab Nebula spectrum 
shows a slight flattening in the spectrum slope towards low energies,
which is consistent with the theoretical expectations
(see~\cite{aa96,dj96}), and which results in a roughly 30\% lower
$\gamma$-ray rate. Thus, the estimate of the trigger performance is
finally $\simeq 3\sigma/\sqrt{hr}$ for a trigger zone of 37 pixels.  
Therefore, 9~hrs observations (37 pixels trigger mode) will provide a detection
of the Crab~Nebula above $\rm \simeq 350~GeV$ with a confidence level
of $\sim 9\sigma$.  

\section{Tests of the trigger threshold} 
Trigger configurations for the lowest possible threshold in the 
{\it topological} trigger mode as well as the stability for
al\-ter\-na\-ting ON and OFF viewing modes\footnote{In the ON and OFF
  viewing mode, the data are taken while tracking alternatively the
  source position (ON) and the sky region chosen for estimate of the
  background (OFF), which is usually placed at $\pm 5^\circ$ away in
  right ascension from the nominal source position.} were tested in
detail for the HEGRA system of IACTs. The trigger threshold was lowered
both by decreasing the threshold of the Discriminator-Monitor-Cards
(DMC), which are used to verify whether a signal in a single PMT
exceeds a certain limit in ph.e. (with a threshold up to 80~mV
equivalent to a range of roughly 0 to 70 ph.e.) and ge\-ne\-ra\-te
a 17~ns pulse supplied into the majority unit for the camera (see for
detailed discussion~\cite{b98}), as well as by increasing the high
voltage (HV) of the PMTs. 19 pixels were included
into the trigger for each camera. The localization of those 19 pixels
were calculated using the altitude and azimuth of the viewing angle
and the actual position of the telescopes in the array. The system was
running in 4-fold coincidence mode. Data runs were taken at 20 degrees
of zenith angle. Table~1 gives a summary of the tested configurations. 
Listed are the gains of the PMTs with respect to the nominal gain
$G_o$, which is calculated from the mean anode currents in PMTs. The
DMC threshold $q^{DMC}_o$ is given in mV, whereas the threshold $q_o$
is measured in photoelectrons\footnote{$q^{DMC}_o$ corresponds to the
  actual setting of the threshold for the DMC, while $q_o$ is the
  value of the threshold converted from mV to ph.e..}. The typical
system event rate, event rate for individual telescope, and the single
pixel rate are summarized in Table~1. $G/G_0$ ratio indicates the
change of the gain after decreasing the threshold, where $G_0$ is a
nominal gain. 

% ----- Table 1 --------------------------------------------------------
\begin{table}[htbp]
\vspace{0.5cm}
\caption{Summary of the test runs taken with the HEGRA system of IACTs
in the topological trigger mode.}
\vspace{5mm}
\begin{center}
%\begin{tabular}{llllll}
\begin{tabular}{cccccc}
\hline
G/G$_0$ & $q^{DMC}_o$  & $q_o$ & System & Telescope & Pixel \\
 & [mV] & [phe] & rate [Hz] & rate [Hz] & rate [Hz] \\
\hline
1.0 &  8 & 6.7 & 0.41 & 1-7 & $3\cdot 10^2$\\
1.53 & 8 & 4.38 & 0.64 & 5-92 & $2\cdot 10^3$\\
\bf{1.45} & \bf{7} & \bf{4.04} & \bf{0.89} & \bf{8-320} & $\bf{4\cdot 10^3}$\\
1.43 & 6 & 3.51 & 0.94 & 37-1.8$\cdot 10^3$ & $10^4$\\
2.07 & 8 & 3.24 & 0.19 & 2.2$\cdot 10^2-7.7\cdot 10^3$ & $1.5\cdot 10^4$ \\
1.89 & 7 & 3.10 & 0.02 & 1.4$\cdot 10^3-8.5\cdot 10^4$ & $3\cdot 10^4$\\
2.40 & 8 & 2.79 & 0.00 & 15$-3.7\cdot 10^3$ & $4\cdot 10^4$ \\
\hline
\end{tabular}
\label{tab:specs}
\end{center}
\vspace{5mm}
\end{table}
% ----------------------------------------------------------------------

The lowest stable trigger threshold before the single pixel rates
exploded was found to be 4.0 photoelectrons, which corresponds to a
PMT gain of 1.45 times the nominal and the threshold of 7~mV. The
system trigger rate was about 0.89~Hz. The nominal threshold was 6.7
photoelectrons, with a DMC threshold of 8~mV, and 1.2~mV/ph.e..
The sta\-bi\-li\-ty of this configuration was tested by changing night
sky background conditions, using a number of 10-minute runs taken with the 
telescopes positioned at the altitude of $80^\circ$ and a\-zi\-muth
$0^\circ$ and brought back to this position at the end of every run.
Within given statistics, for different nights and different night sky
backgrounds, the system trigger rate was always the same for all of
such runs\footnote{Note that, if a PMT current is relatively high due
  to a presence of a bright star in the field of view, this PMT was
  automatically switched off.}.

\section{Summary of Crab Nebula data}
The observations of the Crab Nebula using the {\it to\-po\-lo\-gi\-cal} 
trigger were performed in the so called {\it convergent} data 
taking mode (see \cite{hl99}) instead of the usual parallel mode. 
In the convergent observational mode, the optical axes of
the telescopes intersect at the typical height of the shower maximum
(6-8~km above the sea), instead of being parallel as they are in the
standard parallel mode. The {\it intersection area} in the atmosphere,
which is called the active area (see Figure~4), is substantially
larger in the case of convergent mode. The convergent mode
observations permit to have on average more triggered telescopes per
event for the recorded data than the parallel mode. Moreover, in the
convergent mode, the centers of gravity of the images in the
peripheral telescopes are shifted towards the camera center. 
That gives significantly fewer events with truncated images at 
the edge of the camera and permits an easier selection of the trigger
active areas (see Figure~1). 

% ----- Figure 4 -------------------------------------------------------------
\begin{figure}[t]
\begin{center}
\includegraphics[width=0.8\linewidth]{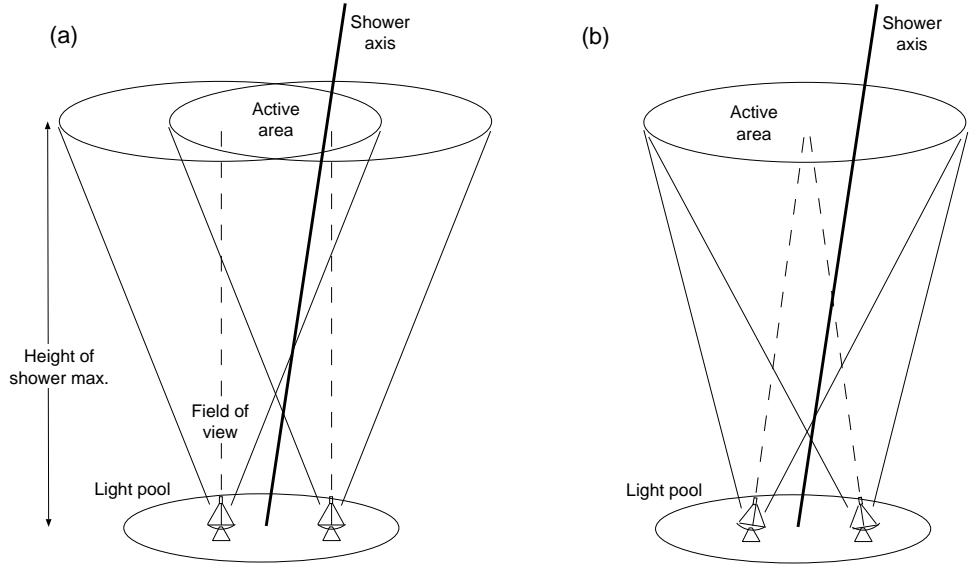}
\end{center}
\caption{\em Schematic view of the {\it parallel} (a) and {\it convergent} (b) modes 
for two telescopes. In the convergent observation mode the optical axes of the 
telescopes intersect at the height of the shower maximum.}
\label{fig4}
\end{figure}
% ----------------------------------------------------------------------------

The Crab Nebula was observed with the topological trigger mode as
described above for a total of 8.7 hours in ON viewing mode and for
about 6.1 in OFF viewing mode. Finally, an area of 37 central camera
pixels was selected for each telescope, with the trigger condition of
2NN/37. The use of 37 pixels instead of the optimal 19 pixels was
motivated by the uncertainty in the tilting angle of system telescopes in
the convergent mode. Even a small displacement of a narrow trigger
zone composed of 19 pixels with respect to the anticipated 
angular area of the $\gamma$-rays images for a certain tilting angle 
of 0.6 degree, will lead to a drastic drop of the $\gamma$-ray rate.
However it is less important for a rather broad trigger zone of 
37 pixels. Table~\ref{tab:data} summarizes the data and the
cor\-re\-spon\-ding rates.  

% ----- Table 2 ------------------------------------------------------------
\begin{table}[htbp]
\caption{Summary of the Crab Nebula data taken with the HEGRA system
  of IACTs in the topological trigger mode.} 
\vspace{0.5cm}
\begin{center}
\begin{tabular}{lcccc}
\hline
\multicolumn{1}{l} {Mode:}  
& \multicolumn{1}{c}{No. of runs}
& \multicolumn{1}{c}{$T_{OBS}$ [hrs]} 
& \multicolumn{1}{c}{Z.A.} 
& \multicolumn{1}{c}{System rate [Hz]} \\
\hline
ON  & 27 & 8.7 & 6$^\circ\div 30^\circ$ & $\sim 1.5 $\\ \hline
OFF & 20 & 6.1 & 6$^\circ\div 30^\circ$ & $\sim 1.5 $ \\ \hline
\end{tabular}
\label{tab:data}
\end{center}
\end{table}
% --------------------------------------------------------------------------

\section{Data analysis}
A number of checks have been performed in order to verify the
$\gamma$-ray signal from the Crab Nebula observed with the HEGRA
system of IACTs using the to\-po\-lo\-gi\-cal trigger mode. First, 
to demonstrate the advancement of the trigger, we have compared the
distributions of the measured image SIZE parameter for the standard
data taking mode and the topological trigger mode. We also compared
the SIZE parameter distributions for data and Monte Carlo events (see
Figure~5). The left plot shows that the content of low energy events
is more prominent for the topological trigger. In addition, the right
panel in Figure~5 shows a good agreement in the SIZE distributions 
between the data and the Monte Carlo simulations. 

% ----- Figure 5- a/b  -------------------------------------------------------------
\begin{figure}[t]
\begin{center}
\includegraphics[width=0.45\linewidth]{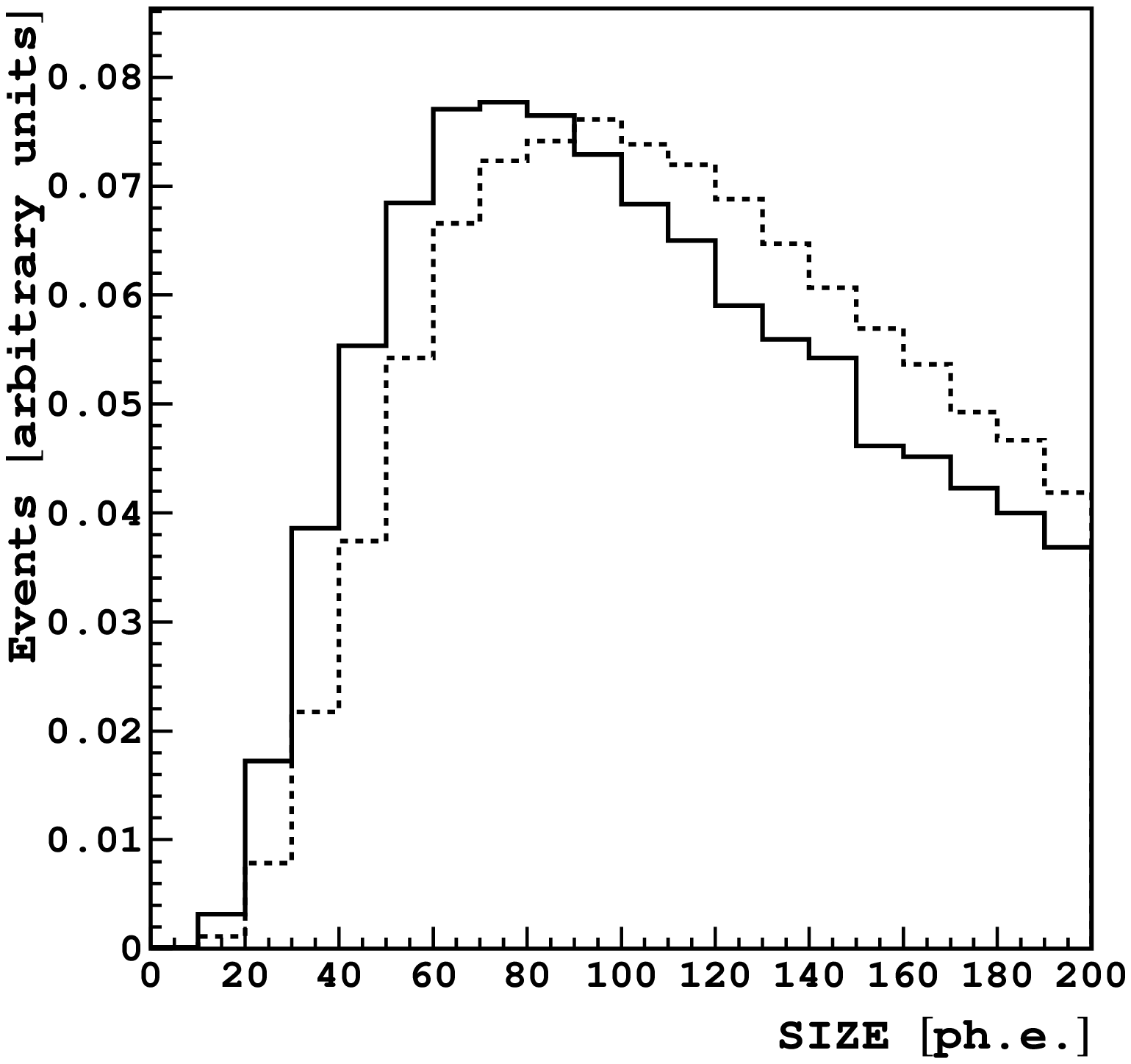}
%\hspace{3.mm}
\includegraphics[width=0.45\linewidth]{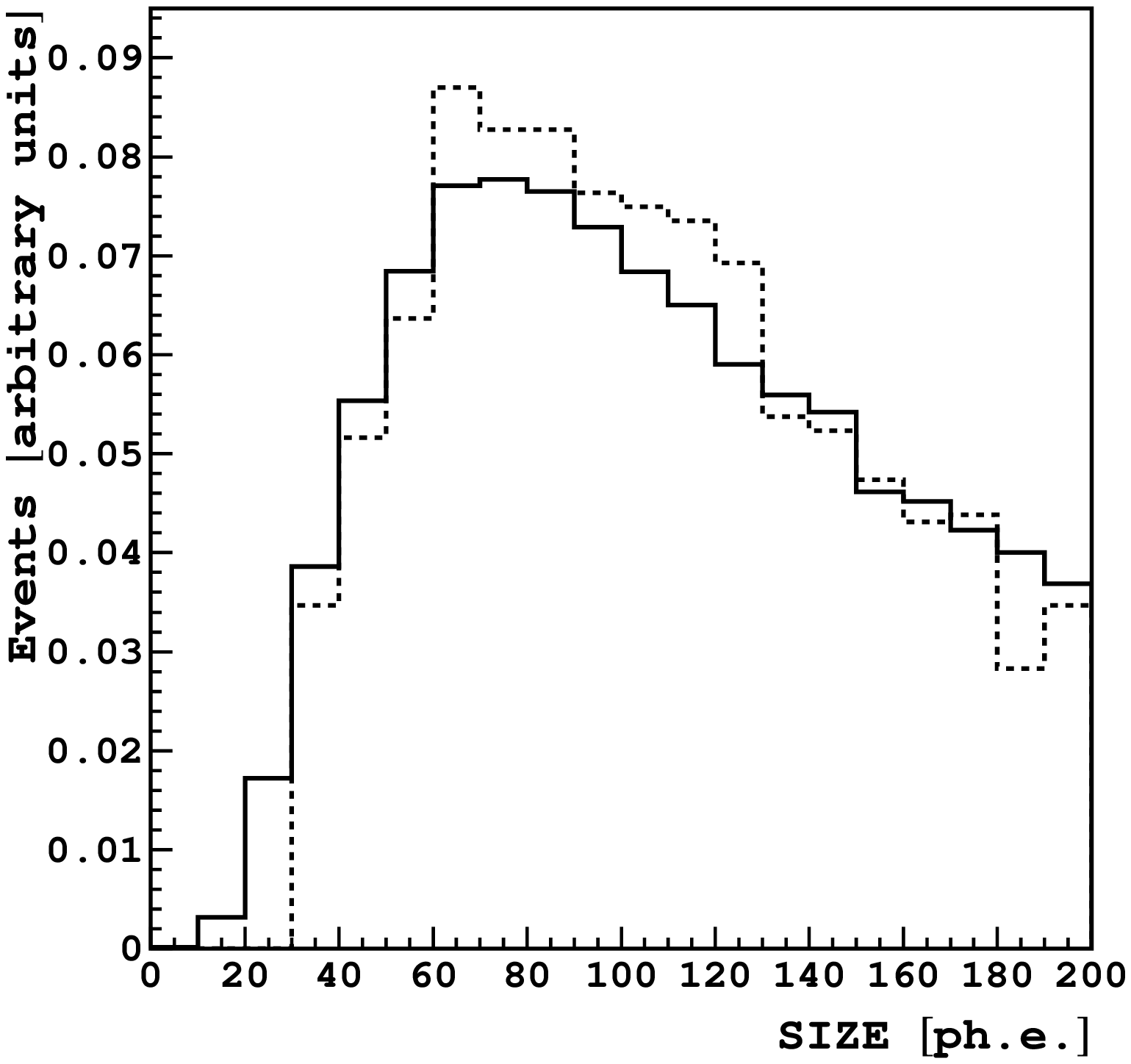}
\end{center}
\caption{\em {\em Left Panel}: Distribution of the SIZE parameter for data 
taken using the topological trigger {\em(solid line)} and normal  
mode {\em(dashed line)}. {\em Right Panel}: Distribution of the SIZE parameter
from data {\em(solid line)} and MC simulations {\em(dashed line)}.}
\label{fig5}
\end{figure}
% ----------------------------------------------------------------------------

Figure~6 shows the distribution of parameter $\langle W \rangle$ for
all recorded excess events, after applying the directional cut of
$\theta^2<0.05\,[deg^2]$. The distribution of excess events peaks at $\langle W
\rangle=0.12^\circ$ and can reproduce well the distribution for the
$\gamma$-rays derived from the Monte Carlo simulations (see for
comparison Figure~3).

% ----- Figure 6 -------------------------------------------------------------
\begin{figure}[htbp]
\begin{center}
\includegraphics[width=0.55\linewidth]{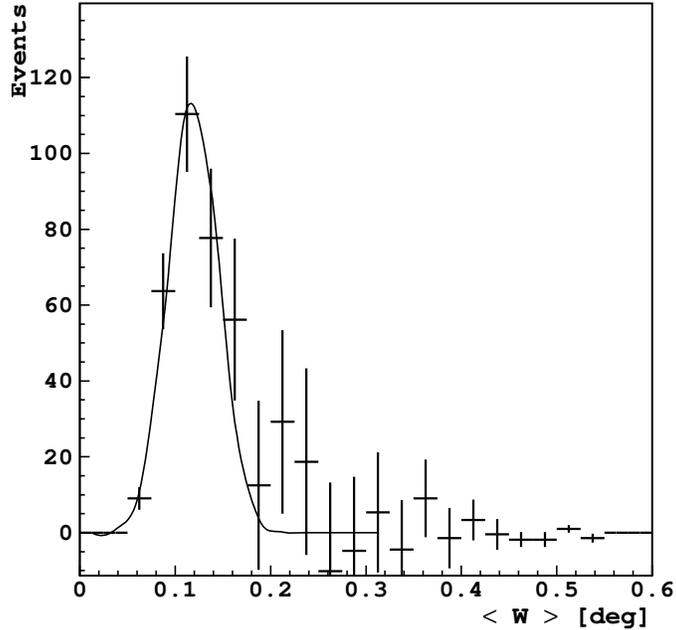}
\end{center}
\caption{\em Distribution of the excess events over the pa\-ra\-me\-ter of weighted WIDTH.}
\label{fig6}
\end{figure}
% ----------------------------------------------------------------------------
   
% ----- Table 3 --------------------------------------------------------------
\begin{table}[bp]
\caption{Acceptances and corresponding significances after applying
  the directional and shape cuts to the Crab Nebula data sample taken
  with the topological trigger mode.} 
\vspace{0.5cm}
\begin{center}
\begin{tabular}{cccccc}\hline
Cuts: & ON & OFF & S/N & S/N/$\sqrt{hr}$ & $\kappa_{\gamma}$ \\ \hline
$\theta^2<0.05\,[deg^2]$; $\langle W \rangle < 0.12^\circ$ & 226 & 72 & 6.1 & 2.1 & 0.53 \\ \hline
\end{tabular}
\label{table3}
\end{center}
\end{table}
% ----------------------------------------------------------------------------

Table~\ref{table3} summarizes the event statistics after applying the
directional and shape cuts. The cut on WIDTH pa\-ra\-me\-ter $\langle W
\rangle < 0.12^\circ$ corresponds to the optimal cut as predicted by
the Monte Carlo simulations. Note that the signal-to-noise ratio (S/N) 
given in Table~\ref{table3} was calculated by using the formula of
Eqn.(17) from \cite{lima}, which takes into account a difference in
exposure times for ON and OFF, whereas the final estimate of the 
system signal-to-noise ratio, derived from Monte Carlo
si\-mu\-la\-tions (Eqn.(4)), was obtained assuming the same
observational time in ON and OFF modes. For the data presented
here, the OFF observational time is by factor of 1.4 less than the
corresponding time for the ON data sample. That explains why
the achieved signal-to-noise ratio is slightly lower than one could
estimate using Eqn.(4). The $\gamma$-ray acceptance, $\kappa_\gamma$,
reported in Table~\ref{table3} was calculated from the real data
comparing the number of excess events before the angular cut (or a
very loose angular cut) with the excess event rate after the actual
orientational cut~\cite{me00}.
 
\section{Energy threshold and flux estimate}
The stereoscopic system permits a straightforward mea\-su\-re\-ment of
shower e\-ner\-gy, based on the re\-con\-struc\-ted impact distance
and the measured SIZE of an image (see \cite{me99}). For the spectrum
studies it is advisable to use a maximal $\gamma-$ray acceptance and
consequently minimize the systematic error related to the cut
efficiency. Here the following cuts were used: $\langle W \rangle <
0.15^\circ$ and $\theta^2<0.05\,[deg^2]$. The same energy reconstruction routine was
used for the Crab Nebula data taken in a standard trigger mode. The
distribution of the recorded $\gamma$-rays over the reconstructed
energy is shown in Figure~7. It is clear that there is a substantial
shift of about 40\% in the position of the peak, which corresponds to
the maximum $\gamma$-ray rate. Note that both data samples correspond
to si\-mi\-lar range of zenith angles. For the topological trigger mode,
the position of the peak is between 200-400~GeV. The energy resolution
of the HEGRA system of IACTs is $\Delta E/E \sim 40\%$ at 0.3~TeV, 
$\Delta E/E \sim 31\%$ at 0.5~TeV, and $\Delta E/E \sim 20\%$ at 1~TeV. As it
is seen in Figure~7, half of the excess events are below the actual HEGRA energy threshold - 500~GeV.
% ----- Figure 7 -------------------------------------------------------------
\begin{figure}[htbp]
\begin{center}
\includegraphics[width=0.7\linewidth]{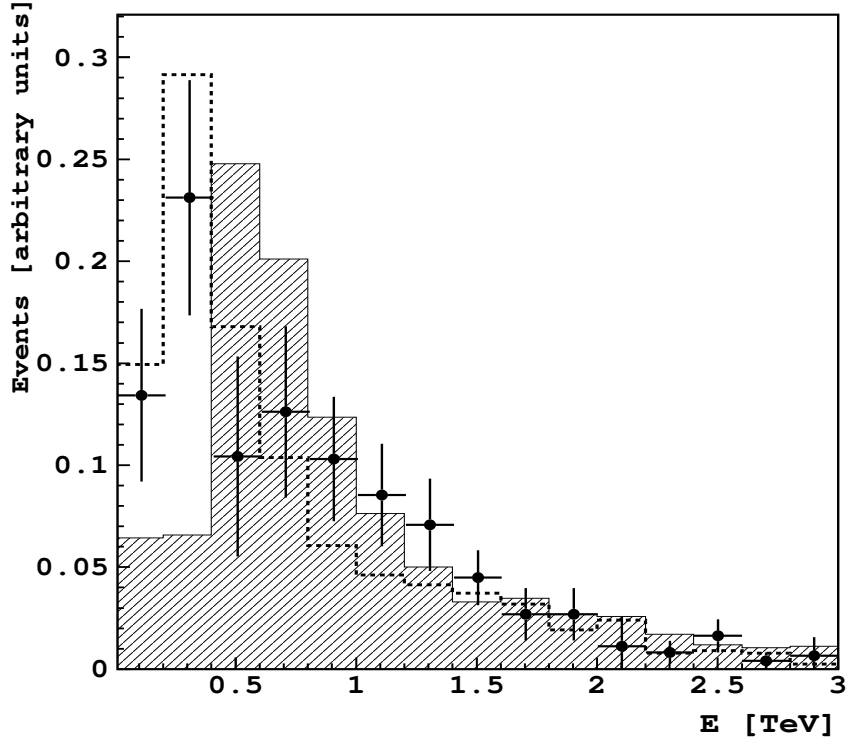}
\end{center}
\caption{\em The energy distribution for the $\gamma$-ray events detected from the
Crab Nebula, from the data taken in topological trigger mode {\em(black
dots)} and standard data taking mode {\em(fill hatched histogram)}. Also
superimposed is the MC prediction {\em(dashed line)} for the topological trigger mode.}
\label{fig7}
\end{figure}
% ----------------------------------------------------------------------------
% ----- Figure 8-9  -------------------------------------------------------------
\begin{figure}[bp]
\begin{center}
\includegraphics[width=0.49\linewidth]{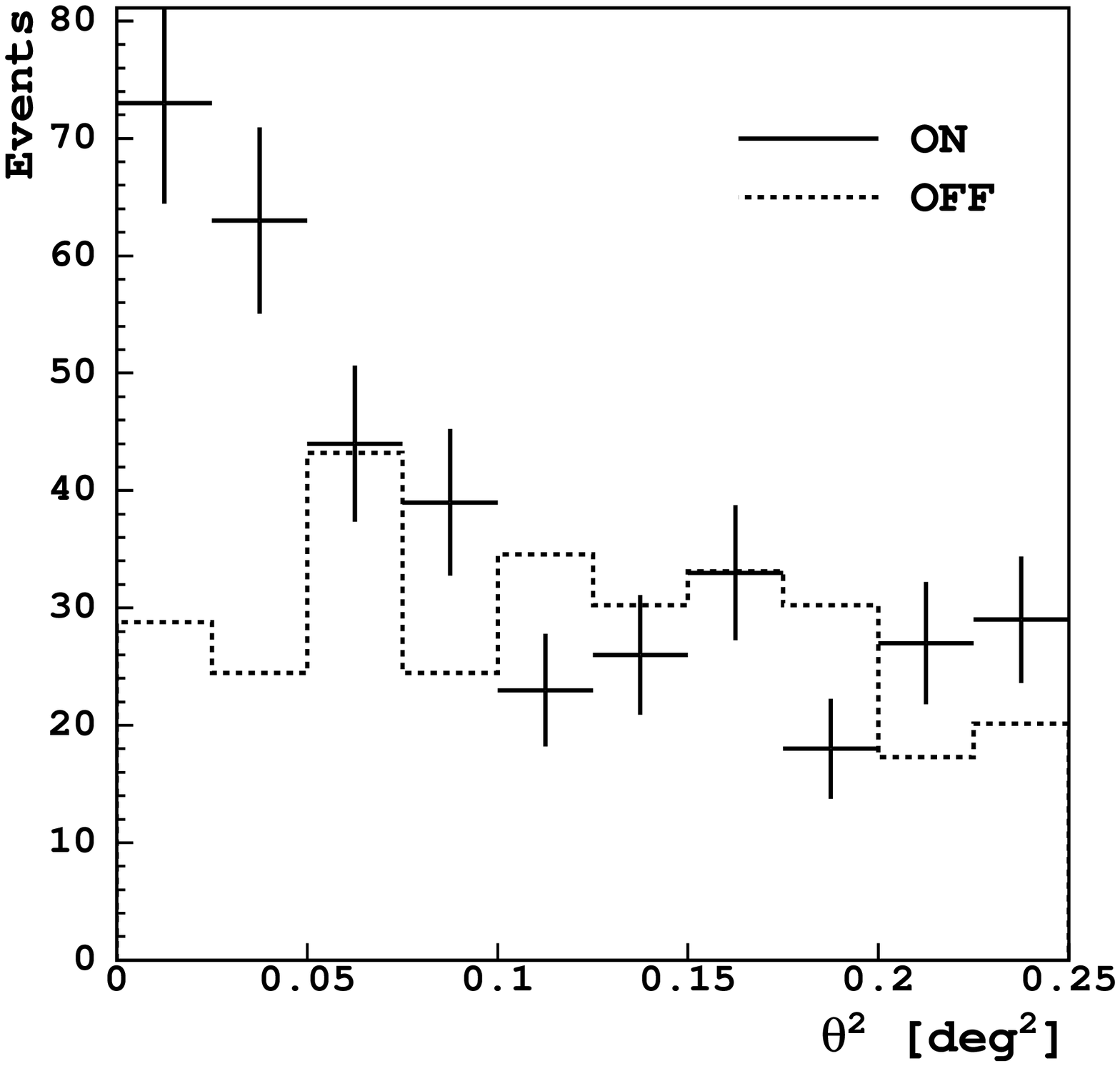}
\hspace{0.3mm}
\includegraphics[width=0.49\linewidth]{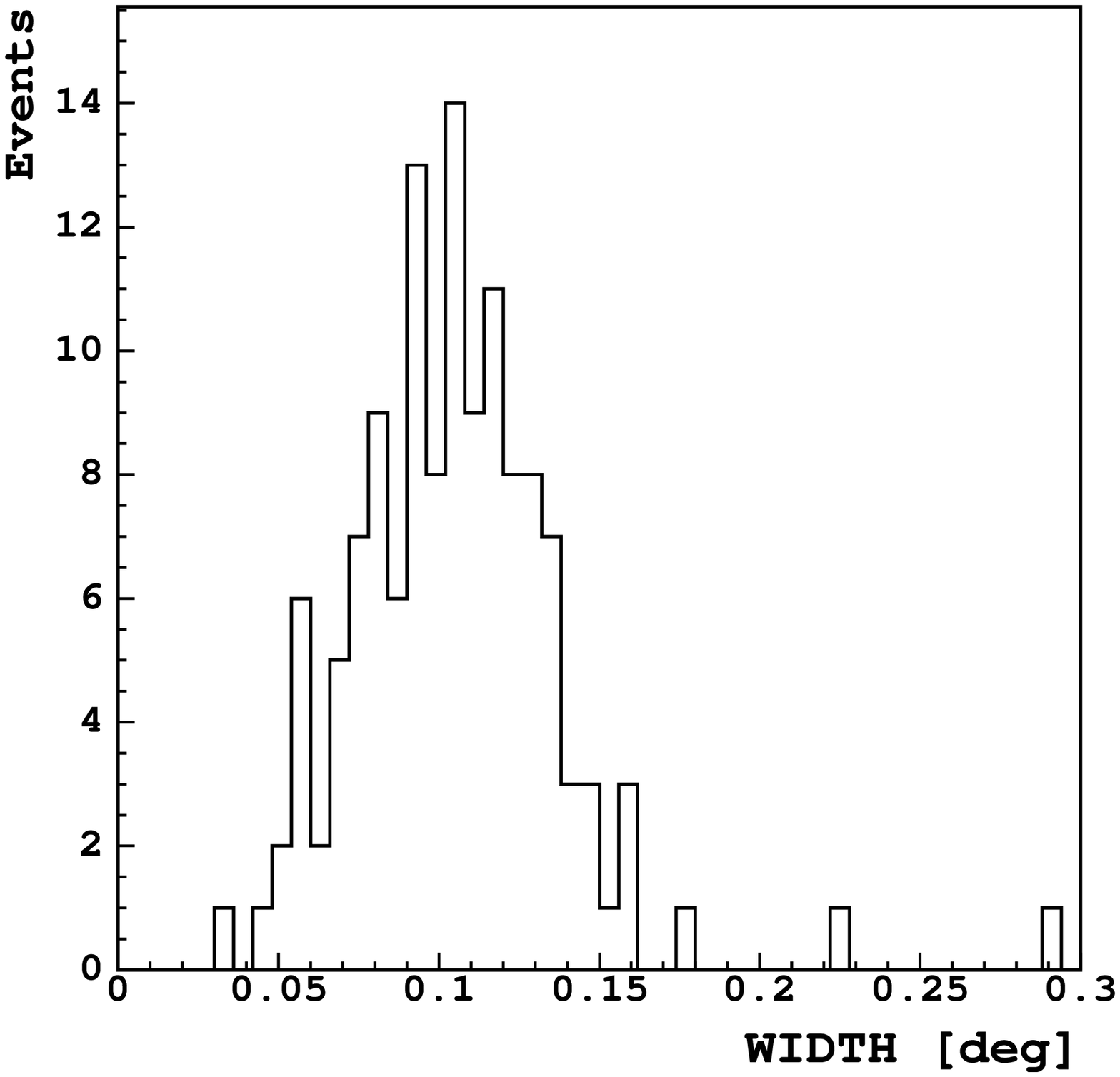}
\end{center}
\caption{\em {\em Left Panel}: Squared angular distance $\theta^2$ of the reconstructed event
directions from the Crab direction with reconstructed energies below 500~GeV 
($\langle W \rangle < 0.12^\circ$). The OFF data have been scaled by a 
factor $\rm T_{ON}/T_{OFF}=1.4$. {\em Right Panel}: WIDTH distribution for CT3 for 
events with $E<500$~GeV.}
\label{fig8_9}
\end{figure}
% ----------------------------------------------------------------------------
Fi\-gu\-re~\ref{fig8_9} shows, respectively, the distributions of the squared 
angular distance $\theta^2$ to the sources ({\em left panel}) and the 
WIDTH parameter for those events below the nominal
energy threshold ({\em right panel}). 

In estimating the $\gamma$-ray flux above the energy threshold, we
use the spectral index of the Crab Nebula power law spectrum as measured by the HEGRA
system of IACTs above 500~GeV~\cite{me00}, which also takes into
account the flattening towards low energies, leaving the normalization factor free: 
\begin{equation}
dJ_\gamma/dE = \rm C(E/1\,TeV)^{-2.47-0.11\,log(E)}\,\, ph\,\,cm^{-2}s^{-1}TeV^{-1}
\end{equation}
For such spectrum one can calculate the total number of expected low energy
$\gamma$-rays 
\begin{equation}
\tilde N_\gamma = \int_{E_o}^{E_1} (dJ_\gamma/dE) S_\gamma(E) k_\gamma(E) dE
\end{equation}
where $S_\gamma$ and $k_\gamma$ are the collection area and acceptance
after the cuts for the $\gamma$-rays, respectively. \\
The boundaries of the integrating regions were selected as $E_o$ = 0~TeV 
and $E_1 = \infty$. By comparing the actual number of recorded
$\gamma$-ray events, $N_\gamma$, with the expected number, $\tilde
N_\gamma$, one can estimate the integral $\gamma$-ray flux as 
\begin{equation}
\noindent J_\gamma(>E_{th}) = (N_\gamma/ \tilde
N_\gamma)\int_{E_{th}}^{E_1}(dJ_\gamma/dE)dE, \,\, E_{th} = \rm 350~GeV
\end{equation}
For the loose analysis cuts the total number of registered $\gamma$-ray 
events is $N_\gamma$=260, thus the estimated $\gamma$-ray flux from the Crab Nebula is 
\begin{equation}
J_\gamma(\rm >350\,GeV) = (8.1\pm0.1\pm0.2)\times 10^{-11}\,\,ph\,cm^{-2}s^{-1}
\end{equation}
Here the estimated statistical and systematic errors are also given. 
The relative variation $\Delta J/ J$ of the integral flux due to the possible 
change of the spectral index is of about 6\%, while the relative 
error on the energy estimation is 15\%.  
The estimate of the integral $\gamma$-ray flux above 350~GeV derived
here is consistent with the measurements by STACEE, CELESTE, WHIPPLE, and CAT
groups (see Figure~9). As it is shown in Figure~9, this estimate of
the integral $\gamma$-ray flux is also consistent with the Inverse
Compton model of the TeV $\gamma$-ray emission. 

% ----- Figure 9 -------------------------------------------------------------
\begin{figure}[htbp]
\begin{center}
\includegraphics[width=0.8\linewidth]{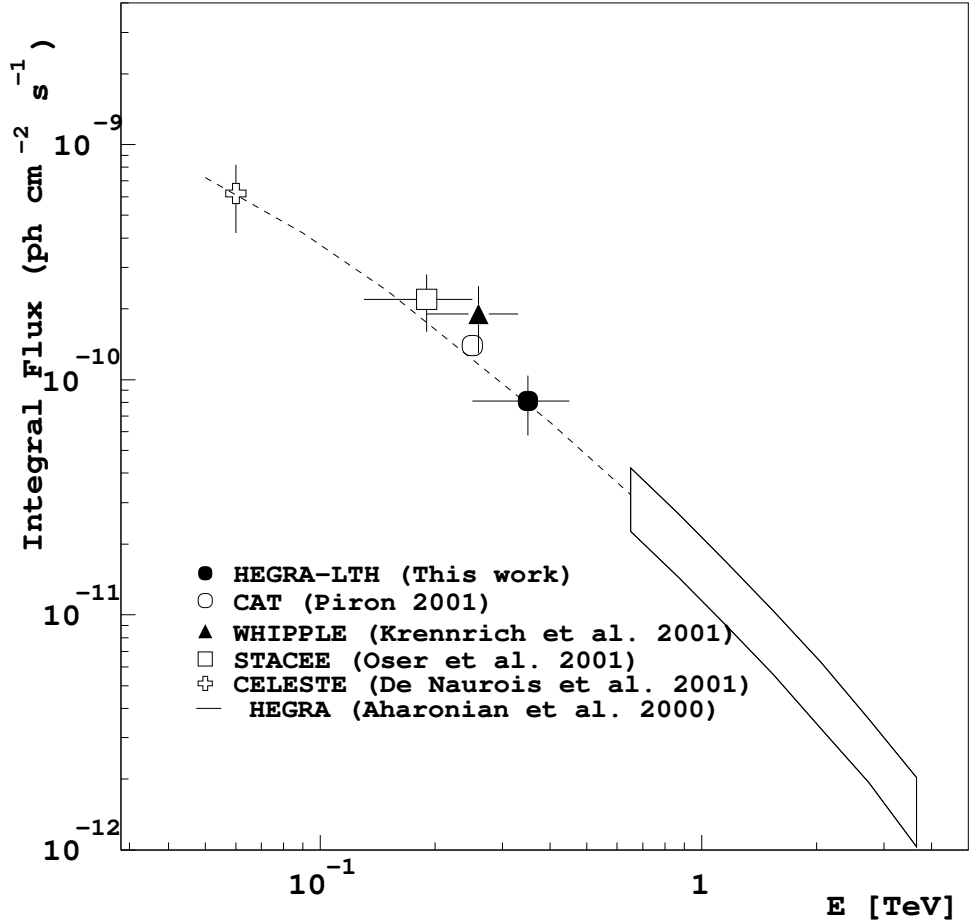}
\end{center}
\caption{\em The integral flux of the $\gamma$-rays from the Crab Nebula measured with the HEGRA 
IACTs array using the topological trigger ({\em black circle}). 
Also shown for comparison are the measurements from other experiments 
and the upper and lower limits of the systematic error estimated for the HEGRA
data. Whipple data were used to calculate the integral flux above 260 GeV
by integrating over the spectrum give in [8] and assuming the systematic
error of 30\%. Dashed curve shows an extrapolation at low energies of the Crab Nebula spectrum as measured by 
HEGRA which also takes into account a flattening in the spectrum slope
towards low energies (see Eq.~(5) and~\cite{me00}).}
\label{fig9}
\end{figure}
% ----------------------------------------------------------------------------

\section{Conclusions}
We performed observations of the Crab Nebula with the HEGRA system of
IACTs using the {\em topological trigger mode}, and demonstrated for
the first time that it is possible to lower the actual energy
threshold of the system by a factor of 1.4 without major hardware
changes and, at the same time, keeping the event rate at a sustainable
level for the currently used DAQ. The Crab Nebula data were taken in
fall of 2000 for a total observational time of 15~hrs to check
the performance of the system in such observational mode. Here we
present the result of the data analysis and give an  
estimate of the integral $\gamma$-ray flux from the Crab Ne\-bu\-la above 
350~GeV. Our estimate of the Crab Nebula flux is consistent with previous
measurements made by STACEE, CELESTE, and CAT groups, and may be
further interpreted as an Inverse Compton TeV emission coming from the 
plerion in the Crab Nebula. 

This technique will be applied in the near future in the observations
with the forthcoming H.E.S.S. ({\it High Energy Stereoscopic System})
system of telescopes, in par\-ti\-cu\-lar to search for sub TeV $\gamma$-ray 
emission from pulsars~\cite{dejager2}. By means of a topological
trigger one can achieve a significant reduction of the energy threshold and
that gives a considerable advantage in such observations,
due to the very steep energy spectrum of the GeV-TeV pulsed emission
as measured by the EGRET detector in the GeV energy range from a
number of such objects. It might be also valuable to apply such technique in search for 
{\it BL Lac objects at rather large red shifts}, which are expected to
have a very steep energy spectrum due to the IR absorption of the
$\gamma$-rays onto the extragalactic background light. 

{\bf Acknowledgments.} The support of the HEGRA ex\-pe\-ri\-ment by the
German Mi\-nis\-try for Re\-search and Tech\-no\-lo\-gy BMBF and by the Spanish
Research Council CICYT is acknowledged. We are grateful to the
Instituto de Astrof\'{\i}sica de Canarias for the use of the site and for
providing excellent working conditions.

\end{document}